# From Risk Prediction Models to Risk Assessment Service: A Formulation of Development Paradigm

Eryu Xia, PhD[1], Yiqin Yu, MSc[1], Enliang Xu, PhD[1], Jing Mei, PhD[1], Wen Sun, PhD[1]
[1] IBM Research – China, Beijing, China

**Abstract**

*Risk assessment services fulfil the task of generating a risk report from personal information and are developed for purposes like disease prognosis, resource utilization prioritization, and informing clinical interventions. A major component of a risk assessment service is a risk prediction model. For a model to be easily integrated into risk assessment services, efforts are needed to design a detailed development roadmap for the intended service at the time of model development. However, methodology for such design is less described. We thus reviewed existing literature and formulated a six-stage risk assessment service development paradigm, from requirements analysis, service development, model validation, pilot study, to iterative service deployment and assessment and refinement. The study aims at providing a prototypic development roadmap with checkpoints for the design of risk assessment services.*

**Introduction**

Risk prediction modeling has been an important topic in clinical informatics research, which estimates the likelihood of a person developing a clinical outcome. Risk prediction models have been developed to predict various outcomes like disease onset in the general population[1–3], development of complications after diagnosis with a certain disease or treatment with a certain drug or procedure[4–6], development of severe outcomes in the critically ill[7–9], hospital readmission after discharge[10], and so on. For example, QRISK[11,12], Pooled Cohort Equation[13], and Framingham risk score[14] have been developed to predict the risk of cardiovascular diseases in the general population. GRACE[15] and TIMI[16] have been developed to predict cardiovascular outcome or mortality in acute coronary syndrome patients[17]. A risk assessment service is developed to provide a risk report based on personal information and has been perceived as helpful in care delivery and clinical decision making[18,19]. While the central component of a risk assessment service is a risk prediction model, which estimates a risk score, a risk report goes beyond by providing risk factors, risk level and intervention. Some existing risk assessment services include: QRISK[11,12] which has been included in 2014 NICE lipid modification guidance as preferred for cardiovascular risk assessment[20], Poole Cohort Equation[13] which has been included in 2013 ACC/AHA guideline[13], and Framingham risk score[14] which has been included in 2009 CCS/Canadian guidelines[21].

Existing literature reported experience and recommendations regarding developing risk assessment services[22–24], but has not proposed an integrated solution. We thus reviewed existing literature and formulated the risk assessment service development paradigm, which contains a roadmap with considerations and actions required in each stage.

Following sections of the article is organized as such: Section 1 reviewed the key elements of a risk prediction model; Section 2 reviewed the key elements of a risk assessment service; Section 3 described our attempt to formulate the risk assessment service development paradigm.

1. **Risk prediction model**

A risk prediction model is an equation or a set of equations estimating the likelihood of developing a clinical outcome based on personal profiles, whose key elements are illustrated in Figure 1 and elaborated below.

**Cohort.** In risk prediction modeling, the cohort can be as general as all people[1–3] and as specific as patients hospitalized due to a certain condition[25], and is usually a group of people likely to develop the clinical outcome. The application of a model should be on a cohort like the one the model was built on to achieve desired performance.

**Predictor.** Predictors are personal characteristics used for risk prediction, including demographics information, vital signs, laboratory test results, family disease history, and so on. For example, commonly used predictors for type 2 diabetes include age, family history of diabetes, body mass index, and waist circumference[1]. Commonly used predictors for cardiovascular disease include smoking, diabetes, hyperlipidemia, hypertension, C-reactive protein, lipoprotein(a), fibrinogen, and homocysteine[26,27]. While including more predictors may increase the prediction performance, it would also increase the burden for data collection. Thus, using less expensive, less invasive, and easily collected predictors, and using fewer predictors may encourage wider application of a model.

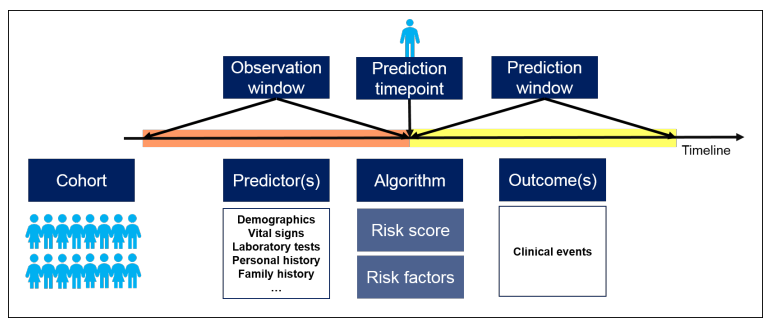

**Figure 1.** Key elements of a risk prediction model.

**Outcome.** Risk prediction models are used to estimate the likelihood of developing a certain outcome. In the clinical domain, the outcomes are often adverse clinical events, including development of disease[1–3], complications[4–6], or severe outcome[7–9], hospital readmission[10], and so on. A model can predict risk for one or multiple outcomes[28]. A previous report suggests that a useful outcome should satisfy the criteria of: (1) undesired to the person experienced it; (2) significant to the health service (cost-effective); (3) preventable; and (4) captured in routine administrative data[24].

**Prediction timepoint.** Prediction timepoint defines when the model is intended to be used. While some models can be widely applied throughout the adulthood, some apply only to people at a certain time point. For example, Rao et al. built a model to predict bleeding complications after patients undergo percutaneous coronary intervention[6].

**Observation window.** Observation window is the period when the predictors are defined and collected.

**Prediction window.** Prediction window is the period when the outcomes are observed. It should be defined based on the outcome. While some outcomes happen shortly after the prediction timepoint, some take years to develop. For example, in predicting hospital readmission, the prediction window ranges from 30 days if reason for the current admission is a severe disease, to 12 months if the reason is a less severe condition or general conditions[10].

**Algorithm.** While the concepts above define the prediction problem, the algorithm solves the defined problem. Two necessary components are algorithms for feature engineering and algorithms to build the risk equation, as elaborated in Box 1.

**Evaluation metrics.** Many evaluation metrics have been proposed to reflect different aspects of the model performance (Box 2). As no model can achieve the best performance in every aspect, how to evaluate a model should be determined based on the aspect of care[29].

---

**Box 1.** Algorithms for modeling

- **Feature engineering.** In risk prediction modeling, feature engineering refers to the process of creating numerical values from predictors to be used in the risk equation.
    - **Pre-processing.** Common pre-processing steps include outlier removal, missing data imputation, standardization, normalization, transformation, encoding, discretization, and so on.
    - **Feature construction and feature selection.** Both serving the purpose of creating meaningful feature sets, they act in different directions. Feature construction aims at generating new features from existing to inject domain knowledge or to increase feature complexity, while feature selection aims at reducing number of features to reduce model complexity and increase model performance.
- **Risk equation.** A risk equation takes engineered features and computes a risk score, which is always achieved through model fitting or a learning process.
    - **Single task.** Most risk equations estimate the risk of a single outcome (a single task).
        - **Classification.** A most common class of risk prediction problems is the binary classification problem

(with vs. without the outcome). Widely used methods include statistical and machine learning approaches like logistic regression[30], support vector machine[31], naïve Bayes[32], Bayesian network[33], and decision trees[34]. Artificial neural networks have also been applied in a few studies[35].
- ◆ **Survival analysis.** Survival analysis is a branch of statistics that analyzes the expected time to an event. Application-wise, we always use survival analysis when the prediction window is long, especially when some data collected have a follow-up period shorter than the prediction window ('censoring'). Widely used survival models include Cox proportional hazard model[15,36] and Weibull accelerated failure time regression model[37].
- ■ **Multiple tasks.** While multiple outcomes can be estimated as multiple separated tasks, there are methods to jointly model multiple outcomes (as multiple tasks). Traditional statistical methods can be extended to multiple tasks by adjusting the model fitting and optimization strategy[38–40]. Artificial neural networks support multiple task modeling intuitively by defining the cost function[41].

---

**Box 2.** Evaluation metrics

- Model prediction performance can be evaluated using various metrics including those from traditional aspects like discrimination and calibration, and from novel aspects like reclassification and clinical usefulness. Here we introduce a few metrics, and more information can be found in previous reports[42].
    - ■ **Discrimination** measures how well a model can discriminate between cases experienced and not experienced the outcome. AUROC (also known as ROC, C-statistics, or C-index) and AUPRC measure the overall performances of a model. Accuracy, sensitivity (recall), specificity, and precision measure model performance at a certain threshold (based on which each sample gets assigned as predicted to 'will have the outcome' or 'will not have the outcome').
    - ■ **Calibration** measures how close the predicted risk score is to the actual probability of developing the outcome. Risk equations from classification methods do not always have good calibration especially in external validation cohort, while those from survival analysis have better calibration. Common metrics include calibration curves and Hosmer-Lemeshow test.
- **Model interpretability** is a measure of how well a model can be understood by human beings.
    - ■ **Model interpretability** reflects how well the mechanism of the model can be interpreted. In a logistic regression model, odds ratio shows both the directionality and the scale of a feature's impact on the outcome. A decision tree model can be easily interpreted as a set of decision rules. Artificial neural networks are notorious for their bad model interpretability due to the black-box nature.
    - ■ **Result interpretability** reflects how well the risk score from a risk model can be explained to a human-understandable degree. Model interpretability leads to, but is not a prerequisite for result interpretability. Methods have been developed to increase result interpretability in complex and black-box models[43,44].

---

2. **Risk assessment service**

While a risk assessment service is based on a model, it has a broader scope. Key elements for a risk assessment service are illustrated in Figure 2. Typically, it starts from collecting personal information as input, to the backend process of feature processing and risk prediction using the algorithm, and finally to the output, a risk report. A report is typically constituted of four components: risk score, risk factors, risk level, and intervention.

**Risk score.** A risk score is a quantitative evaluation of a person's likelihood of developing the outcome in the prediction window and is often a direct or normalized output from the risk prediction model. It can be absolute risk which can be directly interpreted as the probability of developing the outcome, or relative risk which is positively correlated with the absolute risk.

**Risk factor.** A risk factor is an attribute of the assessed individual which has increased his likelihood of developing the outcome[45]. Risk factors overlap with predictors but are different. First, a predictor can be an indicator of increased risk as well as decreased risk, and thus can contribute to as well as protect against the development of the outcome. Second, even when a predictor contributes to development of the outcome, it is a risk factor for a person only if the person has non-optimal value for the predictor. For example, diabetes is a predictor of cardiovascular disease, and for a diabetic person, diabetes is his risk factor for cardiovascular disease. Displaying the risk factors would enable a clear interpretation of the risk score and inform intervention to reduce the risk.

**Risk level.** While risk score is indicative of an individual's own risk, the risk level is indicative of the risk in a comparative manner. Risk level can come from risk stratification which attributes an individual to a risk group of 'low', 'intermediate' or 'high'. Risk stratification can be as simple as stratifying people according to the 'Kaiser pyramid'[24], and can be as complex as a balance considering prevalence of the outcome, cost of mis-stratification, which would be discussed later in this article. Risk level can also be measured as the percentile of the risk score in the cohort or can be the relative risk compared to the average risk.

**Intervention**. While knowing the risk would raise an individual's awareness, knowing the intervention is key to prevent or alleviate the adverse outcome. In an online post, the author raised intervention as the most important recommendations for deploying predictive models, that a model should be deployed with a strategy for intervention and the capability to drive patient care[46]. The intervention can be a specific lifestyle suggestion, initialization of treatment, providing intensive care, or resetting the control target of a clinical indicator.

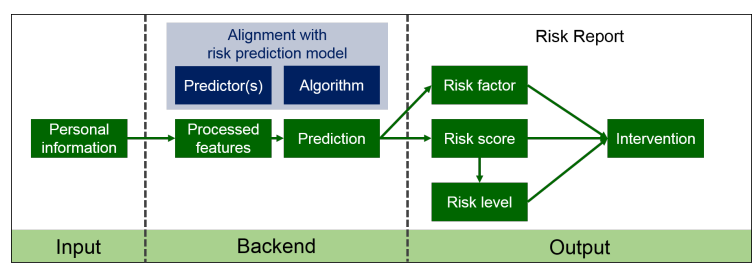

**Figure 2.** Key elements of a risk assessment service.

An example risk assessment service with a report is shown in Figure 3. The JBS3 risk calculator[47] is a tool to help communicate the risk of cardiovascular disease and the benefits of interventions and is designed for doctors and practitioners with their patients. The interface in Figure 3a is used to collect patient information, upon which a risk report could be generated (Figure 3b, c and d). In this example, risk score is the absolute risk of developing a heart attack or stroke in the next 10 years. We take the estimated heart age as the risk level since it is calculated in comparison with others. For intervention, the tool does not explicitly suggest an intervention, but allows exploration of the effects of different interventions.

## 3. A formulation of risk assessment service development paradigm

As a formulation of risk assessment service development paradigm, six stages have been proposed and illustrated in Figure 4: (1) requirements analysis; (2) service development; (3) model validation; (4) pilot study; (5) service deployment; and (6) assessment and refinement. Major efforts in each stage are elaborated below.

### 3.1 Requirements analysis

Requirement analysis should be conducted as the first stage. Here we list a few questions whose answers are prerequisites for developing of a risk assessment service.

**What is the concerned outcome and cohort?** Joint consideration of the outcome and the cohort forms the starting point. Prevalence of the outcome and cohort size should be acquired or estimated to understand and justify the need.

**What is the aim of the assessment?** Different risk assessment services have different aims and focuses. Common aims include: screening from a large population for high-risk persons for further consultation[48], prioritizing high risk patients for a facility or procedure[49,50], providing an individual with his risk factors and the possible interventions[47], and so on. The aim of the assessment is coupled with the deployment scenario and decides the risk prediction modeling process. For example, the aim heavily affects the evaluation metrics of its prediction model. Sensitivity or specificity? Discrimination or calibration? Interpretability or performance? Should the model be using fewer predictors, or should more predictors be included to better uncover the risk factors and suggest interventions?

**What is the deployment scenario?** While the assessment aim is about why the service is needed, the deployment scenario is about where and how the service should be deployed. Common scenarios include: service in emergency rooms to indicate requirements of immediate approach or intensive care[7–9], service for hospitalized patients to inform treatment[50] or prioritize resource utilization[49], service used by general practitioners[11], service in community hospitals for disease screening or disease prevention[49,50], service for personalized disease alert based on self-monitoring data[51,52], and service available to the public. The deployment scenario would inform the risk prediction model development, especially which predictors can be included. It would also enable better analysis of the costs and benefits.

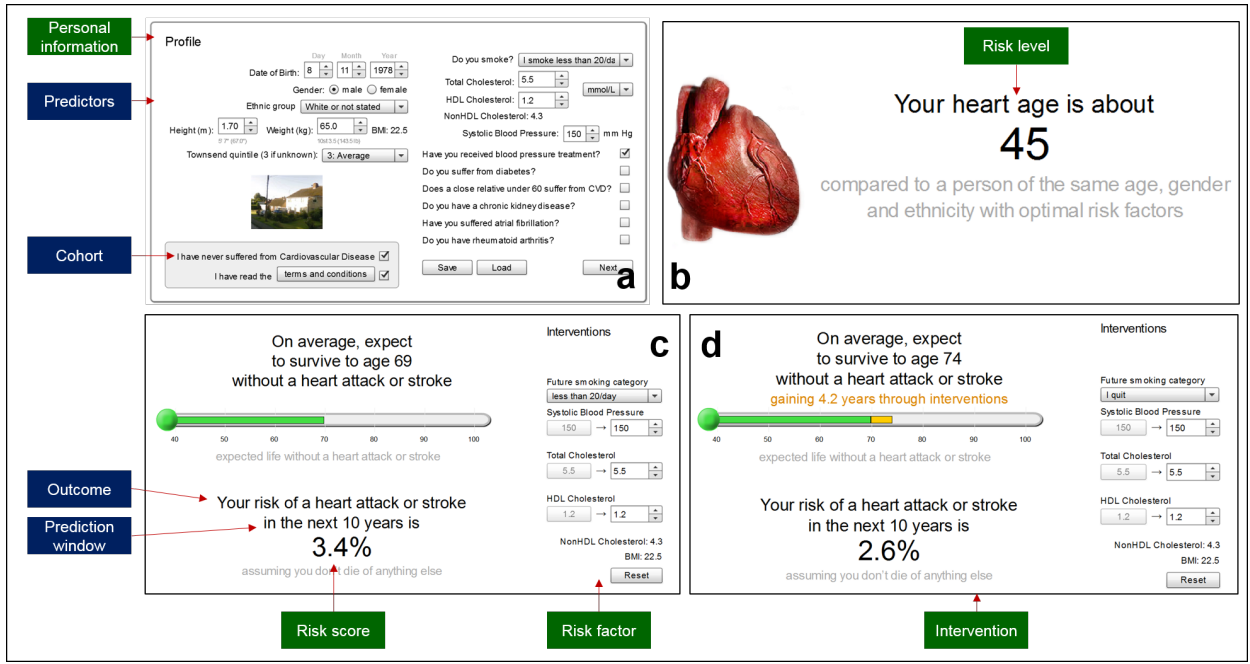

**Figure 3.** Example of a risk assessment service. These are screenshots are from JBS3 risk calculator[47]. After filling in the personal profile (**a**), a heart age would be calculated (**b**), as well as a risk score, risk factors, and potential interventions of the patient (**c**). By adjusting the fields in the 'Interventions' section (here 'Future smoking category' was changed from the current 'less than 20/day' to a prospected 'I quit'), the risk score was updated to reflect the effect of the intended intervention (**d**).

**What should the risk prediction model be like?** For requirements analysis, we should decide, from the risk prediction model perspective, on the model outcome, cohort, prediction window, the evaluation metrics, some of which are already addressed after considering the questions above.

**What are the costs and what are the benefits?** No service comes free of charge. In risk assessment service, the costs are incurred both before and after deployment. Before deployment, the major costs come from data collection and conducting pilot study. After deployment, we should consider direct costs from sustained support for service and collecting predictors, and indirect cost for misleading or inaccurate risk reports. Benefits for persons from the risk report should be considered with the outcome prevalence in the target cohort. The costs and benefits can be monetary or non-monetary and are summarized in Box 3.

### 3.2 Service development

Service development is the stage where we develop the service capable of generating a risk report. Most medical research to develop risk prediction models can fit into this stage. However, actions required in this stage go beyond medical research and may not even include such a research. After requirements analysis, a literature search should be conducted to find any risk prediction model in literature that satisfied the required aim, outcome, cohort, prediction window, and is suitable for the deployment scenario or can be adapted to be such. From there, we would be able to decide on the model source: using existing model from literature, adapting existing model and complementing with local data, or building from scratch using local data. The model source is a major determinant of the data preparation process and the modeling algorithm. If we decide to use an existing model, no data is needed for modeling. If we decide to build on an existing model, we need a small amount of data for model adaptation. If we decide to build a model from scratch, we need to collect a larger amount of data. Algorithms for building models from scratch and algorithms for adapting existing models from literature are introduced in Box 1 and Box 4 respectively. Apart from the model source, the algorithm and modeling process should also take into consideration the available amount of data (to determine model complexity), the aim of the assessment, and the evaluation metrics. Following data preparation, modeling, and model evaluation, we could determine risk level and intervention with the instruction from clinical practitioners, which would then enable generation of a full risk report.

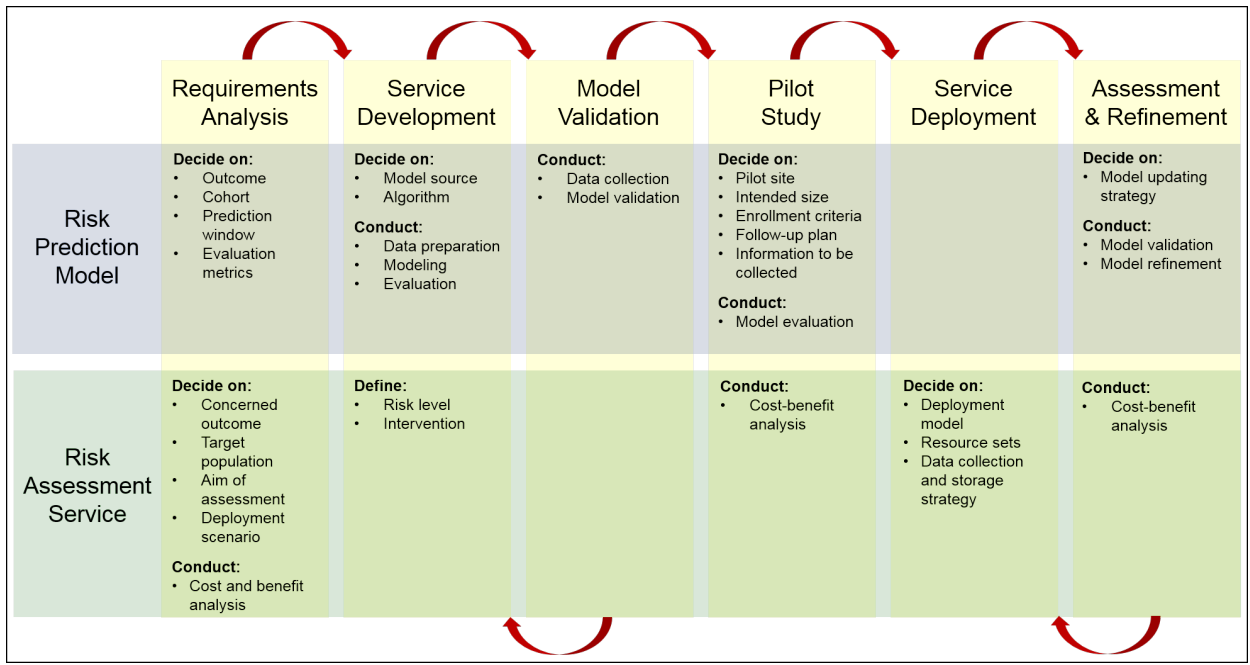

**Figure 4.** A formulation of risk assessment service development paradigm in six stages. In each stage, the key efforts in developing risk prediction model and other efforts are listed in the rows of 'Risk Prediction Model' and 'Risk Assessment Service', respectively.

### 3.3 Model validation

Model validation is the stage where we assess qualitatively and quantitatively how well the service performs based on the evaluation metrics and to which degree it reaches the aim of assessment. In this stage, data needs to be collected, often retrospectively, to validate the model's performance. Unless satisfactory, the process would resume to the service development stage to re-develop the service.

### 3.4 Pilot study

Pilot study is the stage where we implement a small-scale pilot use of the service and evaluate suitability of the service to be deployed in a larger scale. Pilot study differs from model validation from at least two perspectives: (1) while model validation is often conducted as a retrospective study on existing data, pilot study is conducted prospectively with explicit aim and study design; and (2) while model validation evaluates the validity of the generated risk report, pilot study evaluates the service's suitability to be deployed. Pilot study should be conducted as a prospective study with well-designed experiment protocol, including the number and location of pilot sites, the intended study cohort size, enrollment criteria, follow-up plan, and information to be collected at the baseline and upon each follow-up. Apart from the evaluation metrics considered in the service development and model validation stages, we should also conduct cost-benefit analysis (Box 3). Given the results of the pilot study, a decision would be made whether the risk assessment service is ready for large-scale deployment, the process should be resumed to a previous stage, or the development process should be ceased.

### 3.5 Service deployment

The service deployment is the final stage before the service is ready for use. Common service deployment considerations also apply here, like the deployment model (on the cloud, in house or hybrid), resource sets (suitable computing power, memory, and storage), and data collection and storage strategy (centrally or distributively). As healthcare data is especially sensitive, special attention should be paid to guarantee data privacy and confidentiality.

### 3.6 Assessment and refinement

After deployment for a period, data including personal information, actual outcome and overall satisfaction would be collected. Two parts are involved in this stage: assessment and refinement. The same assessment strategy as in the pilot study stage would be applied to evaluate the actual effect of the service. For refinement, we should define the model updating strategy: the criteria whether the model should be refined and the method about how the model should be updated. The model updating methods are described in Box 4, where the principle is to optimize the existing model with newly available data. Assessment and refinement should be conducted iteratively with service deployment.

---

**Box 3.** Cost-benefit analysis

- **Benefit** of a risk assessment service is multi-faceted, which includes but is not limit to[53]:
  - Accurate-prediction benefit:
    - Benefit of cost saving on people need and recommended the intervention.
    - Benefit of cost saving on people not need and not recommended the intervention.
    - Benefit of avoiding intervention's adverse effect on people not needed and not recommended the intervention.
  - Benefit of increasing patient satisfaction by providing suitable intervention[54].
  - Benefit of improved resource planning and allocation based on risk-guided planning and prioritization.

- **Cost** of a risk assessment service is also multi-faceted, which includes but is not limited to[24,53]:
  - Miss-prediction cost:
    - Cost of undesirable outcomes on people need but not recommended the intervention.
    - Cost of wasting money on people not need but recommended the intervention.
    - Cost of intervention's adverse effect on people not need but recommended the intervention.
  - Cost of undesirable outcomes despite recommended the intervention.
  - Cost of intervention.
  - Cost of data collection such as laboratory test cost and follow-up cost.
  - Deployment cost including software and hardware.
  - License fee for using existing proprietary models.
  - Labor costs.

- **Cost-benefit analysis** has been conducted in a variety of studies, which, considering the prevalence of the outcome, performance of the predictive model, intervention strategy, and various costs and benefits, analyzes whether deployment of a risk assessment service is cost-effective and how to harvest the most benefit by selecting the proper intervention strategy. In a study of predicting hospital readmission[55], the authors modeled the net savings or loss (as a combined effect of the total intervention costs and savings) under different intervention strategies. In a recently developed interactive cost-benefit analysis tool[53], total cost can be computed considering the prevalence of the outcome, sensitivity and specificity of the predictive mode, and intervention effectiveness.

---

**Box 4.** Model adaptation and model updating

The aim of **model adaptation** and **model updating** is similar: to build a refined model based on an existing one. While model updating focuses on refining a model to solve the same problem, model adaptation also focuses on adapting a model to be used on a different cohort, at a different prediction timepoint, for a different prediction window, or to predict a related but different outcome. Relevant technical domains are briefly introduced below.

- **Model adaptation**
  - **Transfer learning** is focused on using knowledge gained while solving one problem to help solve a different but related problem. It could reduce requirement of data and learning time.
  - **Domain adaptation** is useful when having a well-performing model trained from a source data and wanting to build a model on data from a different but related distribution.

- **Model updating**
  - **Model recalibration** aims at refitting model parameters without changing the predictors using new data.
  - **Incremental learning** (or dynamic model updating) aims at learning a model to adapt to new data without forgetting its existing model and without retraining.
  - **Build model from scratch** is another option given enough data.

**Discussion and conclusion**

In this study, we reviewed existing literature to leverage previous efforts and formulated a six-stage risk assessment service development paradigm with important actions and considerations in each stage. It aims at providing a prototypic development roadmap with checkpoints for future design of risk assessment services.

Service design is an area in the design field intended for general-purpose service development. There, service design is regarded as an iterative process of four stages: exploration, creation, reflection, and implementation[56]. Our development paradigm is more specific to risk assessment service in the clinical domain and correlates with the general service design methodology in both the iterative nature and the stages. When aligning our proposed stages to the service design framework, requirements analysis corresponds to exploration, service development corresponds to creation, model validation, pilot study, assessment and refinement can be aligned to the reflection stage, and service deployment corresponds to implementation.

We are aware that this roadmap is far from optimal and can be improved from every perspective. A better design of the workflow would both speed up the development process and save resources and efforts. Besides, challenges exist in promoting the use of such risk assessment services, apart from the development itself. This is a multi-party play and the problem is perplexed when involving money and data sensitivity, and thus calls for more exquisite communication and coordination.